\documentclass[useAMS,usenatbib]{mnras}

\usepackage{latexsym}
\usepackage{amsmath}
\usepackage{amssymb}
\usepackage{ascmac}
\usepackage{color}
\usepackage{graphicx,dblfloatfix}
\usepackage{subfigure}

\title[Failure of spherical bodies due to rotation]{Failure modes and conditions of a cohesive, spherical body due to YORP spin-up}
\author[Hirabayashi]{Masatoshi Hirabayashi$^{1}$\thanks{E-mail: masatoshi.hirabayashi@colorado.edu}\\
$^{1}$Aerospace Engineering Sciences, 429 UCB, University of Colorado, Boulder, CO 80309-0429 United States}

\begin{document}

\date{Accepted 2015 August 26. Received 2015 August 26; in original form 2015 August 15}
\pagerange{\pageref{firstpage}--\pageref{lastpage}} \pubyear{2002}
\maketitle
\label{firstpage}

\begin{abstract}
This paper presents transition of the failure mode of a cohesive, spherical body due to YORP spin-up. On the assumption that the distribution of materials in the body is homogeneous, failed regions first appearing in the body at different spin rates are predicted by comparing the yield condition of an elastic stress in the body. It is found that as the spin rate increases, the locations of the failed regions move from the equatorial surface to the central region. To avoid such failure modes, the body should have higher cohesive strength. The results by this model are consistent with those by a plastic finite element model. Then, this model and a two-layered-cohesive model first proposed by Hirabayashi et al. are used to classify possible evolution and disruption of a spherical body. There are three possible pathways to disruption. First, because of a strong structure, failure of the central region is dominant and eventually leads to a breakup into multiple components. Second, a weak surface and a weak interior make the body oblate. Third, a strong internal core prevents the body from failing and only allows surface shedding. This implies that observed failure modes may highly depend on the internal structure of an asteroid, which could provide crucial information for giving constraints on the physical properties. 
\end{abstract}


\section{Introduction}
Recent observations have shown that rotational disruption may be common in our system. P/2013 R3 has broken into multiple components \citep{Jewitt2014}, 62412 has had dust tails from its nucleus \citep{Sheppard2015}, and P/2012 F5 (GIBBS) has ejected dust and fragments \citep{Drahus2015}. They are considered to be candidates that have failed due to fast rotation, implying that rotational disruption may be diverse.  

The YORP effect, which highly depends on the obliquity and shape \citep{Scheeres2007, Statler2009, Cotto2015}, may be one of the primary factors that accelerate/decelerate the spin of an asteroid with a few hundred meters in diameter \citep{Rubincam2000}. We have to mention that small impacts are also capable of changing the spin rate of an asteroid quasi-statically \citep{Marzari2011}. Spin up/down by this effect may be more critical to asteroidal evolution in the solar system \citep{Pravec2007} and in the debris disk around a white dwarf \citep{Veras2014} than ever thought. A remarkable aspect of fast rotating asteroids is that observed asteroids spinning at spin periods shorter than 3.5 hours are all spheroidal (personal communication with Patrick Taylor, 2015). For example, 1999 KW4 Alpha, which is rotating with a spin period of 2.76 hours, is a quasi-spherical body with ridges at its equator \citep{Ostro2006}. The similar features were also seen for the cases of 1950 DA \citep{Busch2007} and 1994 CC Alpha \citep{Brozovic2011}. This may tell us that fast rotating, quasi-spherical asteroids are fairly common in the solar system. It is also important to mention that rotational disruption by the YORP effect would lead to complex dynamical evolution \citep{Jacobson2011}. 

Such analytical and observational evidence of quasi-static spin-up by the YORP effect has been giving rise to the question about evolution and disruption of quasi-spherical objects. Analytical works have contributed to better understanding of failure mechanisms of a uniformly rotating ellipsoid \citep{Dobrovolskis1982, Davidsson2001, Holsapple2001, Holsapple2004, Sharma2013}. Numerical investigations have been capable of dynamically modeling shape reconfigurations by the YORP spin-up \citep{Walsh2008, Walsh2012, Sanchez2012}. These studies gave scientific insights into reshaping processes of ellipsoidal and quasi-spherical objects affected by material friction. Significant progress was made by \cite{Walsh2008} who first succeeded in explaining the formation of equatorial ridges on a quasi-spherical body by using a hard-sphere discrete element code. Also, \cite{Sanchez2012} used a soft-sphere discrete element code to analyze the failure modes of cohesionless ellipsoids having different friction. An ideal shape of an initially spherical body affected by self-gravity and rotation was numerically and analytically investigated by considering the angle of repose \citep{Harris2009, Scheeres2015}. Regardless of these efforts, the reshaping process of a quasi-spherical body is still poorly understood. Specifically, how does the internal structure contribute to evolution and disruption due to quasi-static spin-up? How does the stress field have an effect on failure at severe rotation conditions? How does cohesive strength help an asteroid remain without structural damage? 

The motivation of this study comes from earlier studies about the failure conditions of a quasi-spherical body spinning fast. Assuming a homogeneous structure, \cite{Hirabayashi2015DA} found that 1950 DA may mainly experience failure of its central region, but should not have surface failure at the current spin period, 2.1 hours. This result indicates that landslides may not make equatorial ridges on a quasi-spherical asteroid, but inelastic deformation due to strong compression along the spin axis is a likely driver for it. \cite{Hirabayashi2015SS} confirmed that at high spin rates the central region certainly reaches the yield first and alternatively proposed that to have surface failure and mass shedding driven by it a spherical body must have a strong core. These studies imply that the internal structure plays a prime role in evolution and disruption of a quasi-spherical object. The issue in their works, however, is that they only considered faster spin cases. 

The main objective in this paper is to infer the detailed failure mode of a quasi-spherical object at different spin states and to visualize its evolution and disruption. To do so we consider an asteroid to be perfectly sphere and analyze its failure modes and conditions by applying a continuum mechanics approach. Theoretical and numerical studies are employed here. For the theoretical investigation, analytical solutions of the stress field by linear elasticity are used to predict when and where failure occurs in a spherical body. For the numerical study, a plastic finite element model (FEM) developed by \cite{Hirabayashi2015DA} is employed to see the detailed conditions for failure of the body. 

It should be noted that the analytical study here directly applies the technique by \cite{Dobrovolskis1982}. He introduced a quantity of shear and pressure called ``duress" to investigate the internal structures of the Martian moons, Phobos and Deimos, and several asteroids. According to his study, Phobos should have the maximum shear stress at the central region at the current orbital and rotational configurations, and could only have thrust faults on its surface. He also studied the fate of Phobos, and found that the tidal force will have a primary effect on its failure once it reaches a distance of 6700 km from the center of Mars. We extend his pioneering work to investigate the case in which a spherical object changes its stress field due to quasi-static spin-up. Specifically, we focus on how possible failure modes can shift in the body by considering the Drucker-Prager yield criterion, a pressure-shear dependent yield condition. 

This paper is organized as follows. First, using linear elasticity, we establish a stress field in a spherical body as a function of a spin rate. Second, the yield conditions of the surface region and the central region are analytically investigated to predict the locations of the failed regions that appear first. Third, we use a plastic FEM to evaluate the analytical results. Finally, possible evolution and disruption scenarios of a spherical body are proposed by taking account of a two-layered-cohesive model by \cite{Hirabayashi2015SS}. 

\section{Modeling of a spherical asteroid}
We begin our analysis with modeling of a spherical asteroid that experiences the YORP spin-up. A spherical object is supposed to have a homogeneous structure. Although the model used is the same as \cite{Hirabayashi2015SS}, we do not explicitly consider heterogeneity in the body as our focus is on determining the critical values of material properties with which local regions fail structurally. The density, the radius and the gravitational constant are defined as $\rho$, $R$ and $G$, respectively. The following discussion will use dimensionless quantities. Lengths, body forces, spin rates and stress tensors are normalized by $R$, $\pi \rho G R$, $\sqrt{\pi \rho G}$ and $\pi \rho^2 G R^2$, respectively. The normalized spin rate and radial distance from the center are denoted as $\omega$ and $R_b$, respectively. 

The spherical body is assumed to be rotating along a constant spin axis. Here, this spin axis is lined up in the $z$ axis. The $x$ axis is then given in an arbitrary direction on the equatorial plane, and the $y$ axis is defined to satisfy the orthogonal relation with these axes. This Cartesian coordinate frame is later denoted as $(x,y,z)$. We also use a spherical coordinate system $(R_b, \theta, \psi)$, where $\theta$ is the latitude and $\psi$ is the longitude. 

Material behavior is assumed to be elastic-perfectly plastic -- that is, linearly elastic below the yield condition and no-material hardening on it. To describe such behavior, we use three material properties: Poisson's ratio $\nu$, a friction angle $\phi$ and cohesive strength $Y$. Although we evaluate a stress field based on linear elasticity, since our focus is only on the stress, Young's modulus does not appear in our formulation. Thus, we do not specify this value here. We also consider the latitude $\theta$ a free parameter in the theoretical investigation. 

In this elastic-perfectly plastic model, only elastic strain contributes to stress \citep{Simo1998}. Also, at the yield condition stress does not change, and thus only plastic strain occurs \citep{Chen1988}. On the assumption that plastic deformation is small, this fact guarantees that the yield condition of an elastic stress predicts when and where plastic deformation appears and propagates. To calculate an elastic solution for a spherical body, again, we use the analytical form derived by \cite{Dobrovolskis1982}. 

The present study needs to take into account the fact that plastic deformation is a function of loading paths, which are time-dependent. However, it takes about million years for the YORP effect substantially to change the spin rate of an asteroid \citep{Rubincam2000}, implying that the changes due to time are negligible. A recent breakthrough about the stochastic YORP effect may also provide an interpretation that an asteroid would rather change its spin state stochastically than spin up constantly \citep{Statler2009, Bottke2015}. This intimates that the internal structure may go back and forth between elastic and plastic states. 

\cite{Holsapple2010} who analyzed plastic deformation of a body due to the YORP effect simplified his analysis with the following crucial statement, ``Clearly the possible analyses are vast in parameter space, and are much too complicated to expect closed-form solutions. A much simpler and reasonable approach is to ignore the details of the application and timing of those surface forces, and study the case where the angular momentum is slowly increased in some unspecified but quasi-static way." We follow his assumption to consider the present problem a quasi-static problem. We leave analyses about dynamical evolution of the stress field in an asteroid as our future work. 

We define the terminology about cohesive strength used in this study. The critical cohesive strength is a computed threshold that can prevent an asteroid from experiencing failure. The ``actual" cohesive strength, on the other hand, is an assumed quantity that an asteroid may actually have, which will be constrained by the critical cohesive strength. Also, the spin rate, $\omega = \sqrt{4/3} \sim 1.15$, will be used to describe the condition at which a small particle resting on the surface of a sphere has zero net acceleration \citep{Hirabayashi2015SS}. This simply implies that beyond this spin rate, materials there experience tension and are shed. 

\section{Analytical investigation}
\subsection{Yield condition}
Although \cite{Dobrovolskis1982} defined {\it duress} to express the dependence of materials on pressure and shear, here, we use the Drucker-Prager yield criterion, a smooth function that takes account of shear and pressure: 
\begin{eqnarray}
f = \alpha I_1 + \sqrt{J_2} - s \le 0. \label{Eq:DPcriterion}
\end{eqnarray}
$I_1$ and $J_2$ are the stress invariants: 
\begin{eqnarray}
I_1 &=& \sigma_{1} + \sigma_{2} + \sigma_{3}, \\ 
J_2 &=& \frac{1}{6} \{ (\sigma_1 - \sigma_2)^2 + (\sigma_2 - \sigma_3)^2 + (\sigma_3 - \sigma_1)^2\},
\end{eqnarray}
where $\sigma_i \: (i = 1,2,3)$ is the principal stress component. $\alpha$ and $s$, which are functions of $\phi$ and $Y$, characterize the shape of this yield surface. In general, the radius on the deviatoric stress plane is determined by choosing appropriate $\alpha$ and $s$. To determine these parameters, \cite{Chen1988} compared the radius of the Drucker-Prager criterion with that of the Mohr-Coulomb yield criterion. Here, we use the fact that the stress field in a spherical object is always located at the compression meridian, i.e., $\sigma_1 = \sigma_2 > \sigma_3$ \citep{Hirabayashi2015DA}. Then, $\alpha$ and $s$ are \citep{Chen1988}:
\begin{eqnarray}
\alpha = \frac{2 \sin \phi}{\sqrt{3} (3 - \sin \phi)}, \:\:\: s = \frac{6 Y \cos \phi}{\sqrt{3} (3 - \sin \phi)}. \label{Eq:alpha&s}
\end{eqnarray} 

\subsection{Elastic stress solution in a spherical body}
\cite{Hirabayashi2015SS} used a technique by \cite{Dobrovolskis1982} to provide an elastic stress in a spherical body rotating at spin rate $\omega$. Here, we directly apply their formulation to the present problem. Since the body forces acting on the body are symmetric along the $z$ axis, we consider the stress field at $\psi = 0$:
\begin{eqnarray}
\sigma_{exx} &=& k_1 (1-R_b^2) - k_{14} R_b^2 \sin^2 \theta, \label{Eq:sigxx} \\
\sigma_{eyy} &=& k_1 (1-R_b^2) - k_{13} R_b^2 \cos^2 \theta - k_{14} R_b^2 \sin^2 \theta, \\
\sigma_{ezz} &=& k_9 (1-R_b^2) - k_{14} R_b^2 \cos^2 \theta, \\
\sigma_{exz} &=& k_{14} R_b^2 \cos \theta \sin \theta. \label{Eq:sigxz}
\end{eqnarray}
where
\begin{eqnarray}
k_1 =& \frac{\omega^2}{5} \frac{12 + \nu - 5 \nu^2}{ (1 - \nu) (7 + 5 \nu)} - \frac{2}{15} \frac{3-\nu}{1-\nu}, \label{Eq:k1} \\ 
k_9 =& - \frac{\omega^2}{5} \frac{3 - 6 \nu - 5 \nu^2}{(1 - \nu) (7 + 5 \nu)} - \frac{2}{15} \frac{3-\nu}{1-\nu}, \label{Eq:k9} \\
k_{13} =& - \frac{2 \omega^2}{5} \frac{4 - 3 \nu - 5 \nu^2}{(1 - \nu) (7 + 5 \nu)} + \frac{4}{15} \frac{1-2 \nu}{1-\nu}, \label{Eq:k13} \\
k_{14} =& - \frac{\omega^2}{5} \frac{3 - 6 \nu - 5 \nu^2}{(1 - \nu) (7 + 5 \nu)} + \frac{4}{15} \frac{1-2 \nu}{1-\nu}. \label{Eq:k14}
\end{eqnarray}
The indices of $k$s are chosen to be consistent with those by \cite{Hirabayashi2015SS}. 

\subsection{Characterization of failure conditions}
The yield condition of the stress field is investigated to characterize when and where failure occurs in the body. Specifically, we consider the critical cohesive strength, denoted as $Y^\ast$. From the equal condition of Form (\ref{Eq:DPcriterion}), we obtain $Y^\ast$ as a function of $\omega$, $\phi$, $\theta$, $R_b$ and $\nu$:
\begin{eqnarray}
Y^\ast (\omega, \phi, \theta, R_b, \nu) = \frac{\sqrt{3} (3 - \sin \phi)}{6 \cos \phi} (\alpha I_1 + \sqrt{J_2}). \label{Eq:Ymin}
\end{eqnarray}
As discussed in the previous section, since we assume that a material is elastic-perfectly plastic, if $Y^\ast$ at a given location is higher than the {\it actual} cohesive strength, that location reaches the yield. Thus, giving the contour curves of $Y^\ast$ predicts the failed regions in the body. A negative value of $Y^\ast$ indicates that the locations having it do not need cohesive strength to avoid the yield. 

We analytically explore how $Y^\ast$ evolves as $\omega$ changes. To do so we compute $Y^\ast$s at the body's origin and on the surface, later denoted as $Y^\ast_c$ and $Y^\ast_s$, respectively. At the origin, where $R_b=0$, the stress invariants are:
\begin{eqnarray}
I_1 &=& 2 k_1 + k_9, \label{Eq:I1c} \\
J_2 &=& \frac{1}{3} (k_1 - k_9)^2. \label{Eq:J2c}
\end{eqnarray}
Then, we obtain $Y^\ast_c$:
\begin{eqnarray}
Y^\ast_c (\omega, \phi, \nu) &=& \frac{3 - \sin \phi}{6 \cos \phi} \left[ \frac{3+2 \nu}{7+5 \nu} \omega^2 \right. \nonumber \\ 
&& \left. + \frac{\sqrt{3}}{5} \alpha \frac{(3 - \nu) (-2 +\omega^2)}{1 - \nu} \right]. \label{Eq:YminCenter}
\end{eqnarray}
Needless to say, $Y^\ast_c$ is independent of $\theta$. When $Y^\ast_c = 0$, the spin rate should be:
\begin{eqnarray}
\omega^{\ast 2}_c = \frac{2}{1 + \frac{5}{\sqrt{3} \alpha} \frac{1 - \nu}{3 - \nu} \frac{3 + 2 \nu}{7 + 5\nu}}.  
\end{eqnarray}
This condition is a key parameter that tells us when the central region needs cohesive strength to avoid failure. This parameter indicates a lower threshold showing when a spherical body becomes oblate. 

On the surface, where $R_b=1$, on the other hand, the stress invariants are:
\begin{eqnarray}
I_1 &=& - (k_{13}+k_{14}) \cos^2 \theta - 2 k_{14} \sin^{2} \theta, \label{Eq:I1s} \\
J_2 &=& \frac{1}{3} \{ (k_{13}^2 - k_{13} k_{14} + k_{14}^2) \cos^4 \theta + k_{14}^2 \sin^4 \theta \nonumber \\
&& + k_{14} (k_{13} + k_{13}) \sin^2 \theta \cos^2 \theta \}. \label{Eq:J2s}
\end{eqnarray}
Then, $Y^\ast_s$ and the spin rate at which $Y^\ast_s = 0$, denoted as $\omega^\ast_s$, are obtained as functions of $\theta$ by substituting these equations into Equation (\ref{Eq:Ymin}); however, since the expression is too long to write down here, we only mention that the derivation is trivial. 

\subsection{Results}
Hereafter, following the fact that the failure conditions are not strong functions of Poisson's ratio \citep{Holsapple2008, Hirabayashi2014Kleo}, we fix Poisson's ratio at 0.25, which allows for compressibility of the volume. For $\phi$, we consider a range of $25^\circ - 45^\circ$ for a typical geological material \citep{Lambe1969}. 

\begin{figure*}
	\begin{center}
		\subfigure[]{
         	\label{Fig:yphi25}	
		\includegraphics[width=\columnwidth]{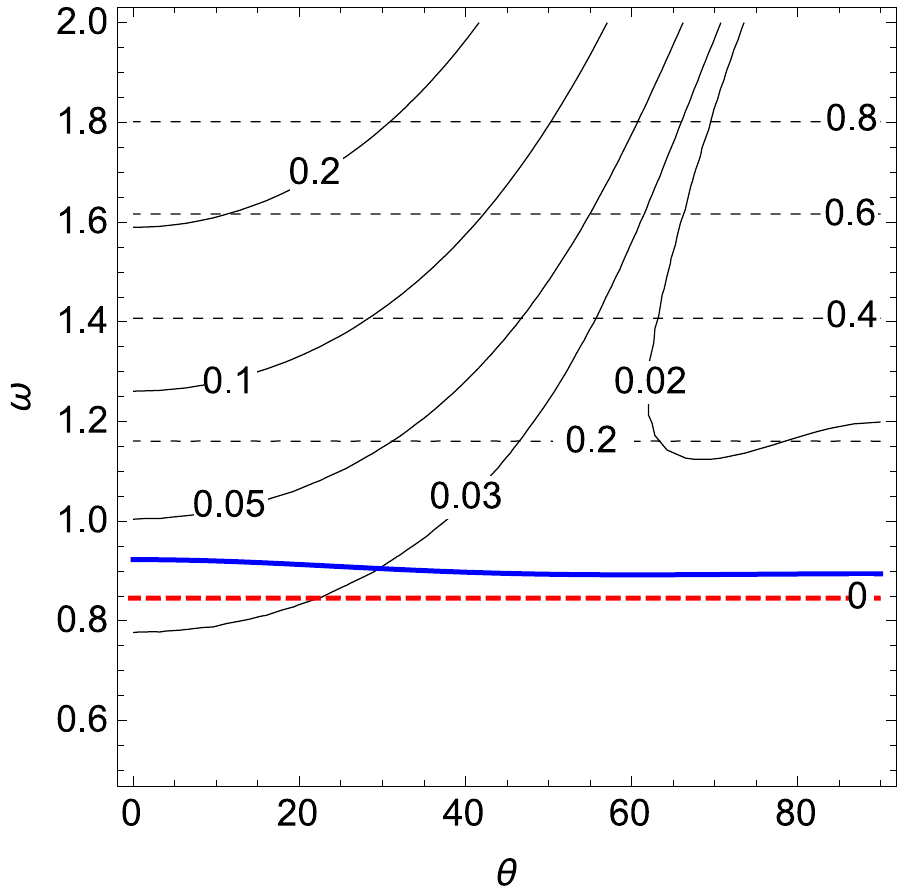}
          	}
		\subfigure[]{
         	\label{Fig:yphi35}	
		\includegraphics[width=\columnwidth]{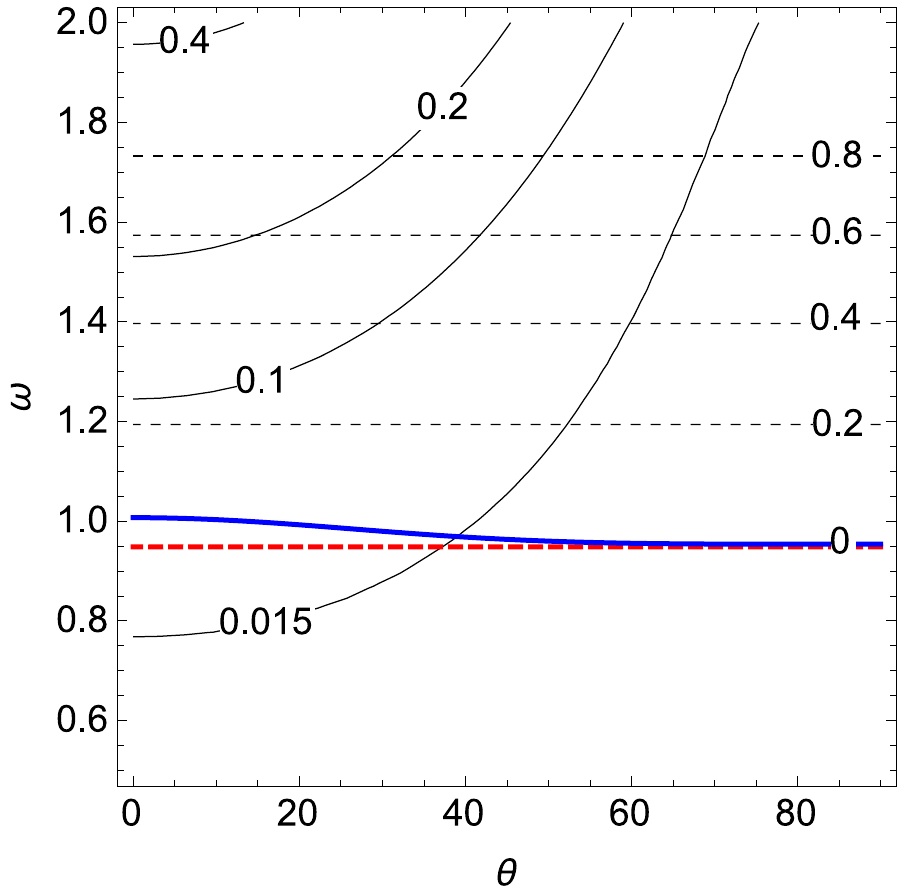}
          	} \\
		\subfigure[]{
         	\label{Fig:yphi45}	
		\includegraphics[width=\columnwidth]{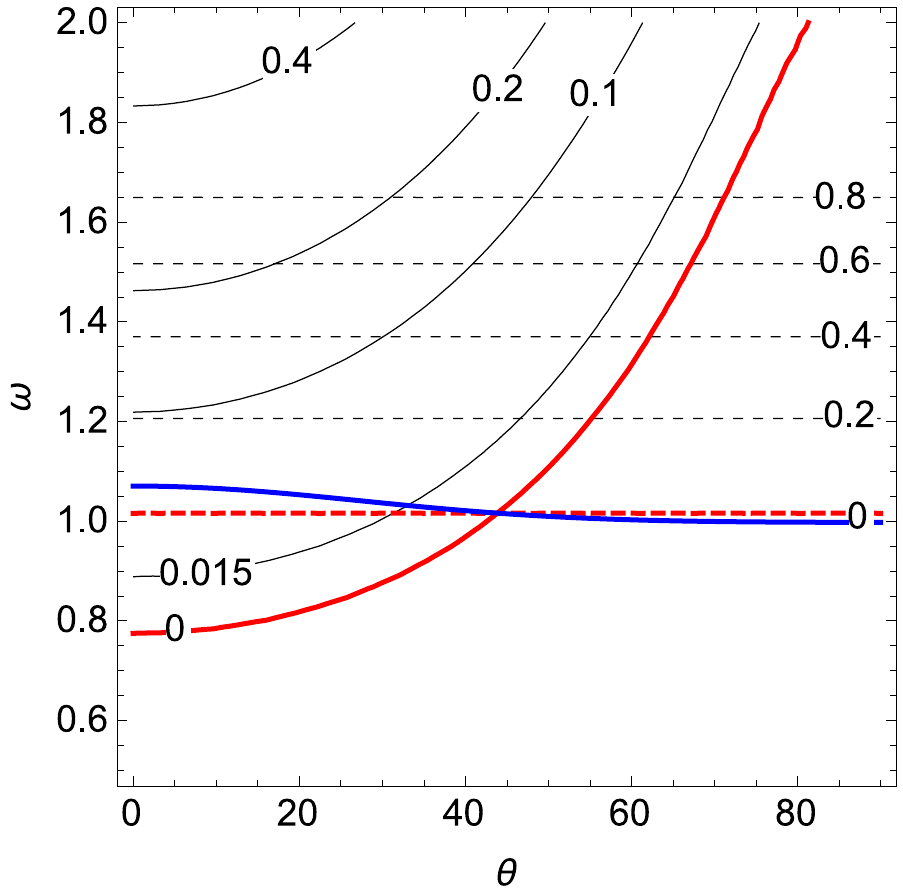}
          	}
	\caption{Contour curves of $Y^\ast$. The solid and dashed curves indicate $Y^\ast_s$ and $Y^\ast_c$, respectively. Figures \ref{Fig:yphi25}, \ref{Fig:yphi35} and \ref{Fig:yphi45} describe the cases of $\phi=25^\circ$, $35^\circ$ and $45^\circ$, respectively. The red curves indicate the conditions of zero-cohesive strength for both cases. The blue lines give $\omega$ at which $Y^\ast_c = Y^\ast_s$. The color version is available online.}
	\label{Fig:yphi}
	\end{center}
\end{figure*}

Figure \ref{Fig:yphi} shows the contour plots of $Y^\ast_c$ (the dashed curves) and $Y^\ast_s$ (the solid curves) in a $\theta-\omega$ space. Figures \ref{Fig:yphi25}, \ref{Fig:yphi35} and \ref{Fig:yphi45} display the cases of $\phi=25^\circ$, $35^\circ$ and $45^\circ$, respectively. The red solid and dashed curves provide $\omega^\ast_s$ and $\omega^\ast_c$, respectively. The blue curves show the condition at which $Y^\ast_c = Y^\ast_s$, which means that if the spin rate is higher (lower), the central region is more (less) sensitive to failure than the surface. The fact that the blue line and $\omega^\ast_c$ are close to one another shows how rapidly $Y^\ast_c$ increases. Thus, $\omega^\ast_c$ can be considered to a parameter describing when the central region starts to fail structurally. For the cases of $\phi=25^\circ$ and $35^\circ$, $Y^\ast_s$ is always positive. The analytical expression of $Y^\ast_{s}$ shows that it is always positive if $1 - 12 \alpha^2 > 0$, i.e., $\phi < \sin^{-1} (3/5) \sim 36.87^\circ$. 

\begin{figure*}
  \centering
  \includegraphics[width=4.8in]{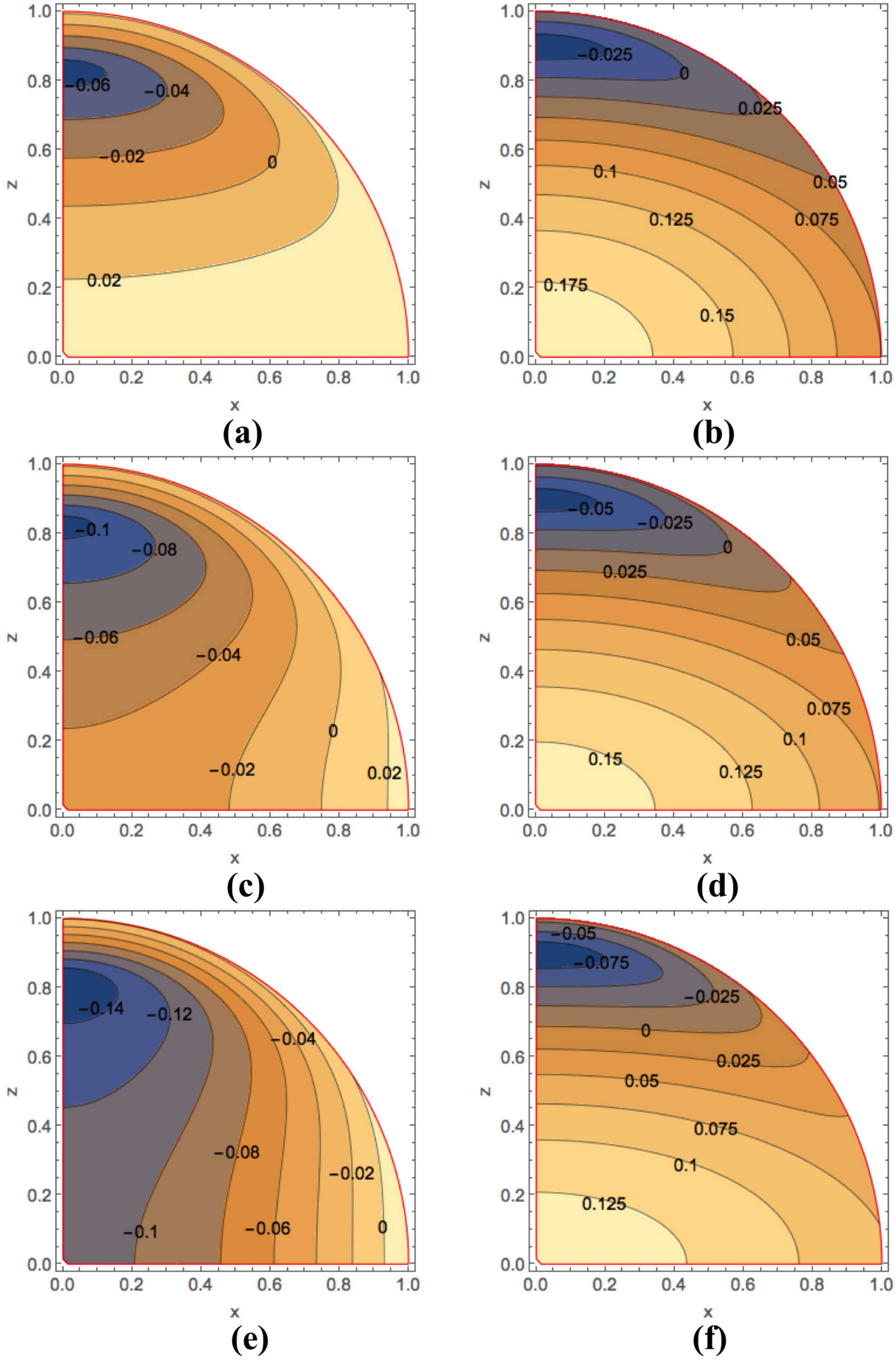}
  \caption{Distribution of $Y^\ast$ across the quadrantal cross section along the $x$ and $z$ axes. The elastic solutions are used to provide this distribution. The left and right columns describe spin rates of 0.9 and 1.15, respectively. The first, second and third rows indicate friction angles of $25^\circ$, $35^\circ$ and $45^\circ$, respectively. The color version is available online.}
  \label{Fig:AnalyticalModel}
\end{figure*}

We plot the distributions of $Y^\ast$ on the $x-z$ plane for different spin rates and friction angles in Figure \ref{Fig:AnalyticalModel}, which are analogous to Figure 4 in \cite{Dobrovolskis1982} except that he displayed the maximum shear. The left and right columns indicate spin rates of 0.9 and 1.15, respectively. The 0.9 spin rate is chosen based on $\omega^\ast_c$. Figures \ref{Fig:AnalyticalModel}(a) and \ref{Fig:AnalyticalModel}(b), \ref{Fig:AnalyticalModel}(c) and \ref{Fig:AnalyticalModel}(d), and \ref{Fig:AnalyticalModel}(e) and \ref{Fig:AnalyticalModel}(f) display the cases of friction angles of $25^\circ$, $35^\circ$, and $45^\circ$, respectively. 

For the case of a friction angle of $25^\circ$, $\omega^\ast_c = 0.85$ and $Y^\ast_s > 0$ everywhere. At a spin rate of 0.9, if the {\it actual} cohesive strength, later simply denoted as $Y$, is less than 0.02, a region enclosing the surface and the central region becomes the most sensitive to failure (Figure \ref{Fig:AnalyticalModel}(a)). We note that if the spin rate is less than $\omega^\ast_c$, the central region should be stable. At a spin rate of 1.15, however, since $Y^\ast$ becomes maximum at the central region, if $Y$ is less than, for example, 0.175, the central region where $Y^\ast > 0.175$ should fail structurally (Figure \ref{Fig:AnalyticalModel}(b)). 

For the case of a friction angle of $35^\circ$, since $\omega^\ast_c = 0.95$ and $Y^\ast_s$ is positive everywhere, the blue curve is always higher than, but close to, $\omega^\ast_c$ (Figure \ref{Fig:yphi35}). Contrast to the $25^\circ$ friction angle case, at a spin rate of 0.9, since $Y^\ast$ is negative at the central region, if $Y$ is less 0.02, only the equatorial surface should fail (Figure \ref{Fig:AnalyticalModel}(c)). At a spin rate of 1.15, on the other hand, and $Y^\ast$ is the highest at the central region (Figure \ref{Fig:AnalyticalModel}(d)). For the case of a friction angle of $45^\circ$, $\omega^\ast_c$ for this case is 1.01. The results show the same features as the $35^\circ$ friction angle case; however, the region where $Y^\ast > 0$  only appears  at the equatorial edges. $\omega^\ast_c$s for these cases are always lower than $\omega \sim 1.15$. 

These cases describe that at high spin rates, a spherical body requires higher cohesive strength to avoid failure of the central region. This corresponds to earlier studies \citep{Hirabayashi2015DA, Hirabayashi2015SS}. Plastic deformation should be compressive along the spin axis and tensile on the equatorial plane. However, if the spin rate is lower than $\omega^\ast_c$, the equatorial surface is the most sensitive, which is one of the findings in this study. A variation of the friction angle causes a change of $Y^\ast$; the higher the friction angle is, the smaller $Y^\ast$ is. As a result, $\omega^\ast_c$ increases because a higher friction angle can prevent the body from failing at faster spin rates. For any cases, however, $Y^\ast$ of the equatorial surface ($\theta = 0^\circ$) becomes positive at slower spin rates than that of the central region; the failure regions move from the equatorial surface to the central region as the spin rate increases (Figure \ref{Fig:yphi}). This result indicates that the failure mode is not a strong function of a friction angle. It should be noted that this results from linear elasticity, which does not take into account plastic flow yet. Since plastic flow can make wider regions fail, the actual failure region should be wider than the current prediction \citep{Chen1988}. More precise propagation of plastic deformation in the body has to be discussed based on the flow rules, and we leave this discussion as our future work. 

\section{Numerical investigation by plastic FEM}
We use a plastic FEM developed by \cite{Hirabayashi2015DA} to evaluate the analytical results above. The code is developed by using commercial finite element program ANSYS 15.03. In this model, elastic behavior is linear, while plastic behavior is perfectly plastic and follows an associated flow rule, which defines that plastic deformation always occurs in the normal direction to the yield surface. It is also assumed that plastic deformation is small. We create a 10-node tetrahedral FEM mesh for a spherical body. Six node displacements are fixed so that the rotational and translational motions are cancelled out. Similar to the analytical model, we consider materials in the body to be homogenous. The yield condition is the Drucker-Prager yield criterion, which is described in Equations (\ref{Eq:DPcriterion}) through (\ref{Eq:alpha&s}).

\begin{figure*}
  \centering
  \includegraphics[width=6in]{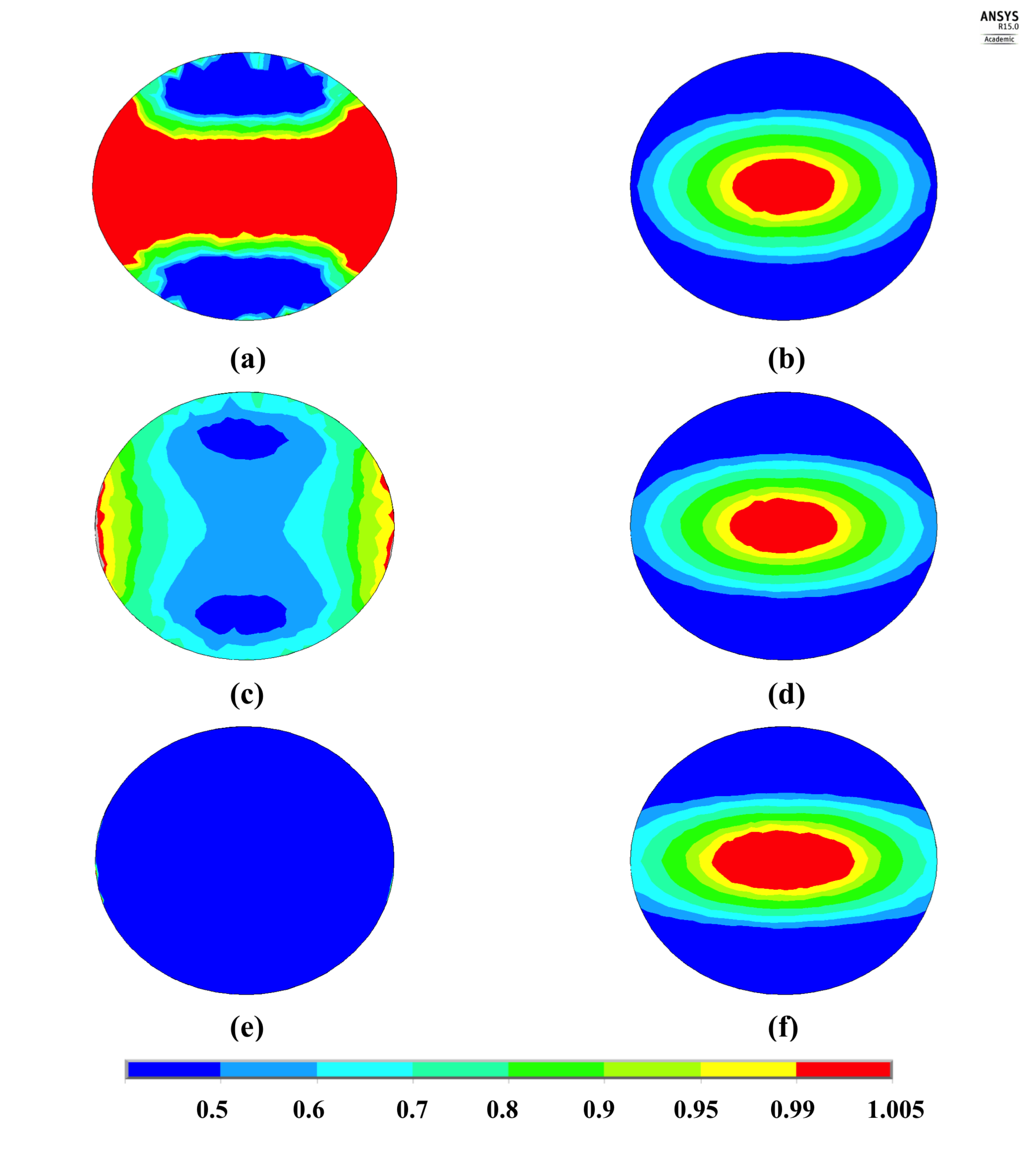}
  \caption{Failure regions on the cross section. The horizontal and vertical axes are identical to the $x$ and $z$ axes, respectively. The plastic FEM simulations are conduced to obtain these failure regions. The simulation conditions displayed here are the same as Figure \ref{Fig:AnalyticalModel}. The color version is available online.}
  \label{Fig:NumericalModel}
\end{figure*}

Figure \ref{Fig:NumericalModel} shows the results by the plastic FEM for the same cases as Figure \ref{Fig:AnalyticalModel}. The contours indicate the stress ratio, a ratio of the current stress to the yield stress \citep{ANSYSThr}; if this ratio is 1, the region should experience plastic deformation, which is given in red. To induce the plastic deformation seen in this figure we choose the {\it actual} cohesive strength for each case. 

For the case of $\phi = 25^\circ$, we choose $Y=0.02$ at $\omega = 0.9$ to induce failure in both the central and surface regions (Figure \ref{Fig:NumericalModel}(a)) and $Y=0.175$ at $\omega = 1.15$ to produce that in the central region (Figure \ref{Fig:NumericalModel}(b)). For the case of $\phi = 35^\circ$, we input $Y=0.02$ at $\omega = 0.9$ to trigger failure on the equatorial surface (Figure \ref{Fig:AnalyticalModel}(c)), and $Y = 0.15$ at $\omega = 1.15$ to make the central region fail (Figure \ref{Fig:AnalyticalModel}(d)). Last, for the case of $\phi = 45^\circ$, $Y=0.02$ at $\omega = 0.9$, and $Y = 0.125$ at $\omega = 1.15$. The case for $\omega = 0.9$ does not have any failed regions in the body, while that for $\omega = 1.15$ causes the similar failure mode appearing in the central region. We confirm that the sizes of the failure regions by the plastic FEM are always wider than those by the analytical study. Regardless of this, all the results by both of the techniques are fairly consistent. We have to note that Figure \ref{Fig:NumericalModel} shows the deformed shapes after loading. 

\section{Disruption scenario of a spherical body}
The present analysis provided insights into evolution of a cohesive, spherical body due to the YORP spin-up. Using the results obtained above, this section discusses possible evolution and disruption scenarios of a spherical body by taking into account a two-layered-cohesive model by \cite{Hirabayashi2015SS}. On the assumption that the bulk density is homogeneous, their model consists of a spherical core which has higher cohesive strength and a surface shell which possesses low cohesive strength. They concluded that if there is a strong internal core, a possible disruption mode is surface shedding.  

\begin{figure*}
  \centering
  \includegraphics[width=6.5in]{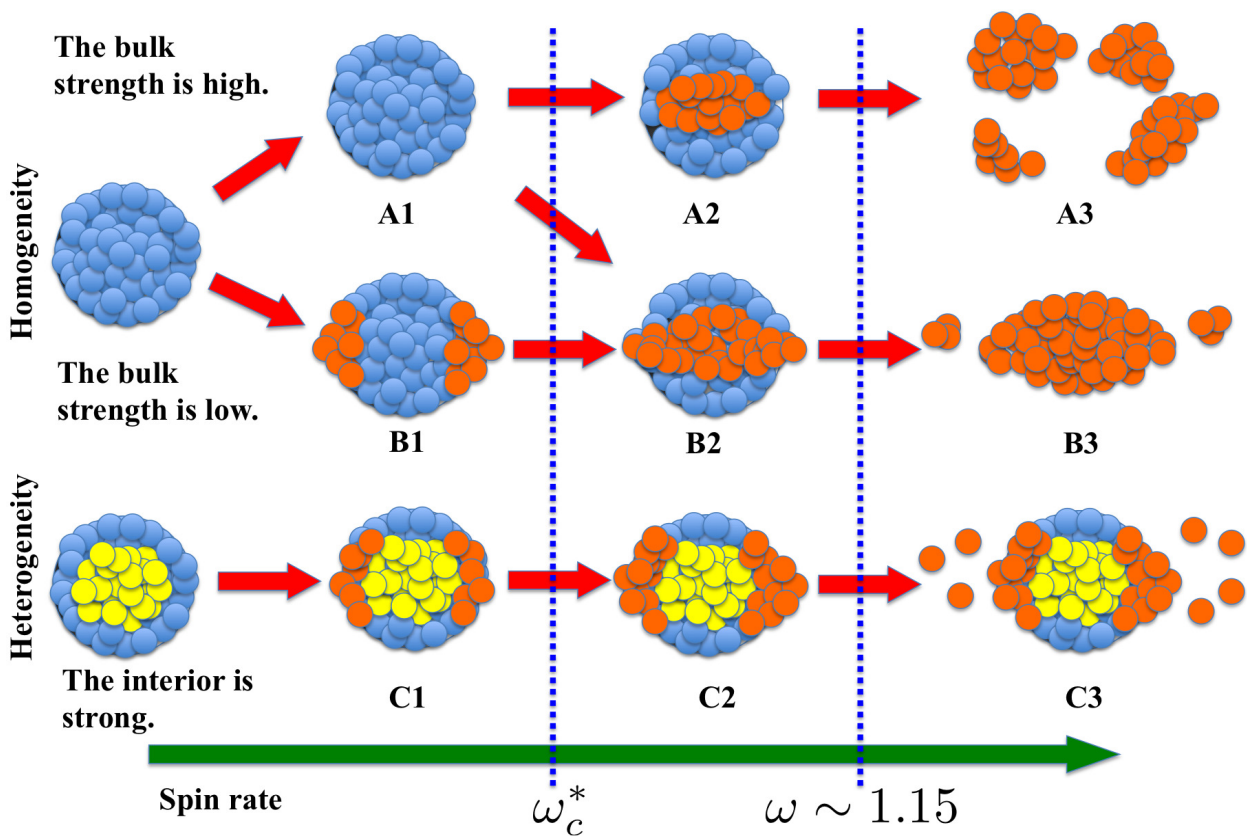}
  \caption{Evolution and disruption scenarios of a spherical body. The blue and yellow markers show elastic components, and the orange markers describe failed ones. The blue components are structurally weaker than the yellow components. As the spin rate increases, the disruption mode evolves differently due to the internal structure. The color version is available online.}
  \label{Fig:diagram}
\end{figure*}

Figure \ref{Fig:diagram} indicates evolution paths of a spherical body. Along every path (from the left to the right), the {\it actual} cohesion is considered to be constant. $\omega^\ast_c$ and $\omega \sim 1.15$ are used to characterize the stages of the evolution. A spherical body with homogeneous structure would have two pathways to its terminal state. If the {\it actual} cohesive strength is high enough for the body to support its structure at a slow spin rate, there are no failed regions in the body (A1). As the spin rate becomes higher than $\omega^\ast_c$, the central region then reaches the yield first (A2). At this spin rate, although the failed region becomes wider, since the body can still sustain itself, disruption does not occur yet. If the {\it actual} cohesive strength is not high enough, stage A1 can result in stage B2, which is discussed later. The body at stage A2 reaches its terminal state at a spin rate higher than $\omega \sim 1.15$ (A3). The {\it actual} cohesive strength is so high that the body can spin fast; however, once the structure fails structurally at a high spin rate, large components immediately fly away. This disruption mode is a breakup of the body into multiple components \citep{Hirabayashi2015SS}. 

If the {\it actual} cohesive strength is low at a low spin rate, the equatorial surface fails first (B1). Stage B2 is a phase in which both surface and central regions fail at a spin rate higher than $\omega^\ast_c$, leading to propagation of plastic deformation over the entire equatorial plane. This stage can come from two stages, A1 and B1. At this spin rate, since the equatorial plane has already failed, the body becomes oblate due to vertical compression \citep{Hirabayashi2015DA}. Once the spin rate is higher than $\omega \sim 1.15$, the shape is more and more oblate (B3). The oblateness may depend on a friction angle because higher friction can prevent local elements from deforming inelastically (personal communication with Paul S\'anchez, 2015). However, since plastic deformation highly depends on loading paths \citep{Chen1988}, if the body stays at critical conditions for a long time, the oblateness may grow critically even for high friction angle cases. Since the surface has failed already, a large amount of mass may be ejected from the edges \citep{Sanchez2012}. The prediction of this mode can be seen from Figures \ref{Fig:AnalyticalModel} and \ref{Fig:NumericalModel}; if the {\it actual} cohesive strength is small, failure widely spreads out across the surface and central regions, leading to vertical compression \citep{Hirabayashi2015DA} and mass ejection. 

If the body has a strong core and weak surface, it has a different pathway to its terminal state \citep{Hirabayashi2015SS}. At a spin rate of less than $\omega^\ast_c$, a failed region appears on the equatorial surface first (C1). This mode should be identical to stage B1. However, as the spin period becomes higher than $\omega^\ast_c$, since the central region is strong enough, the surface region suffers failure (C2). A considerable failure mode at this stage is a landslide because the central region has high cohesive strength to resist failure, and the surface region should fail due to its low cohesive strength. The shape at stage C2 may look similar to that at stage B2. However, since the formation process is different, it may be possible to distinguish between stages B2 and C2 by observing different surface morphology. The shape at stage B2 may not have the morphology created by landslides, while that at stage C2 may have. Once the spin rate is above $\omega \sim 1.15$, only the surface mass is shed without large deformation of the original body (C3). This disruption can be distinguished from stage B3 because the amount of ejected mass and the final shape can be controlled by the size of the internal core. If the core is large, inelastic deformation is limited and only a limited amount of mass can be ejected from the surface. 

\section{Discussion}
The present analysis shows that the disruption modes of asteroids may be dependent on the internal structure. On the assumption that reported active asteroids are spherical, the evolution diagram in Figure \ref{Fig:diagram} shows that an asteroid having a breakup, such as P/2013 R3, should be structurally strong enough to fail at a high spin rate (A3). 62412 and P/2012 F5 (GIBBS) may be between stages B3 and C3 because large components were not seen in their dust tails \citep{Sheppard2015, Drahus2015}. Again, we emphasize that this is given only under the assumption that they are spherical at this point. In fact, they would be elongated rather than spherical. Further investigation is necessary to explore elongated cases. Another interesting asteroid is 311P/PANSTARRS, formerly P/2013 P5, which has ejected a small amount of mass from the body \citep{Jewitt2013, Jewitt2015}. This mode is consistent with stage C3. Also, the size of dust particles was only a few hundred microns. This gives rise to a possible scenario that the spin period of this body is not too fast, so ejected boulders may not have enough kinetic energy to escape from the main body. Note that our study does not explain the episodic event of this object. Again, further investigation is necessary to explain this unique event.

It is interesting to consider possible formation scenarios for a quasi-spherical object with equatorial ridges. Either stage B2 or stage C2 may represent the observed shapes such as 1999 KW4 Alpha \citep{Ostro2006}, 1950 DA \citep{Busch2007} and 2008 EV5 \citep{Busch2011}. To have these stages, they should go through one of the following processes: A1 $\rightarrow$ B2, B1 $\rightarrow$ B2 and C1 $\rightarrow$ C2. To go through process B1 $\rightarrow$ B2, a homogeneous body has to have small cohesive strength. Based on Figure \ref{Fig:AnalyticalModel}(c), for a friction angle of $35^\circ$, a spherical body should have an {\it actual} cohesive strength of less than 0.02 at a spin rate of 0.9. For these asteroids, we find that the {\it actual} cohesive strengths of 1999 KW4 Alpha, 1950 DA and 2008 EV5 should be less than 8 Pa, 5 Pa and 0.5 Pa, respectively. For 2008 EV5, we assume its bulk density to be 2000 kg/m$^3$. Note that these values are quite small compared to the reported cohesive strength of 1950 DA, which is $> 75-85$ Pa \citep{Hirabayashi2015DA}, and that of P/2013 R3, which is $40 - 210$ Pa \citep{Hirabayashi2014R3}. 

We also investigate $\omega^\ast_c$ for these asteroids. The central region begins to deform at this spin state and becomes more oblate as the spin rate increases. Assuming a friction angle to be $35^\circ$, we obtain this boundary as $\sim 3$ hours for 1999 KW4 Alpha, 1950 DA and 2008 EV5. Since the spin periods of 1999 KW4 Alpha and 1950 DA are 2.7 hours and 2.1 hours \citep{Ostro2006, Busch2007}, respectively, these asteroids can become more oblate due to failure of their central regions at the current spin states. On the other hand, the spin state of 2008 EV5, which is 3.8 hours \citep{Busch2011}, is below this lower threshold, implying that the body may structurally relax. However, because of its shape, it might have been spinning at a spin period of shorter than 3 hours. 

Finally, we mention the earlier studies about the shape evolution of a cohesionless, spherical body. \cite{Harris2009} developed a numerical model that computes an ideal shape of a spherical body by considering the angle of repose. Extending their study analytically, \cite{Scheeres2015} found that in the two-dimensional model, once failure occurs on the surface, the material that comes off of the mid-latitudes should leave behind material at the angle of repose for the regolith. This is different from our results that showed that the equatorial surface should fail first at a slow spin rate. The difference between our work and his may come from our computational settings. Specifically, our boundary condition may be different from his. Our surface condition is based on a zero-traction condition \citep{Hirabayashi2015SS}, while he defined the surface stress based on the force balance on the surface. These are not necessarily the same. We will investigate the connection between our work and his in the future. 

\section{Conclusion}
This paper explored how failed regions in a spherical body evolve as the spin rate changes and deduced its terminal state based on our results and the earlier works by \cite{Hirabayashi2015DA} and by \cite{Hirabayashi2015SS}. The main contributions are as follows:

\begin{itemize}
\item Prediction of the failed regions by the elastic model is consistent with the results by a plastic FEM. 

\item The surface region at the equator needs higher cohesive strength to avoid failure at a low spin rate, while the central region must have much higher cohesive strength as the spin rate increases. 

\item There is a spin rate condition, given as $\omega^\ast_c$, at which the central region needs cohesive strength to avoid failure. If the spin rate is higher than this condition, a spherical body may become oblate. 

\item $\omega^\ast_c$ increases as the friction angle becomes higher; the friction angle changes the failure conditions. However, even at different friction angles, the failed regions consistently move from the equatorial surface to the central region as the spin rate increases; the failure modes are not strong functions of it.

\item Three possible disruptions were discussed in this study. For a homogeneous body, a strong structure allows the body to remain at a higher spin rate and then causes it to have a breakup into multiple components. A weak structure leads to failure on the equatorial surface at a slow spin rate, and then has failure propagation to the central region. As a result, at its terminal state, the body becomes oblate. For a body with a strong core, a possible failure mode is only surface shedding. 

\item The formation of the equatorial ridges on a spherical body depends on cohesive strength. If the surface region is very weak, the ridge should be made only by downslope movement, which mainly occurs at lower latitudes; otherwise, the main driver for it is inelastic deformation along the spin axis. 
\end{itemize}

\section*{Acknowledgements}
\addcontentsline{toc}{section}{Acknowledgements}

The author wishes to thank Dr. Patrick Taylor and Dr. Paul S\'anchez for useful comments on this study and Dr. Tony Dobrovolskis for his careful and constructive reviews that improved this paper. The license of ANSYS 15.03 is owned by Aerospace Engineering Sciences, The University of Colorado at Boulder. 


\bsp	

\label{lastpage}
\end{document}